\documentclass[twocolumn,noshowpacs,preprintnumbers,amsmath,amssymb]{revtex4-2}
\usepackage{graphicx}
\usepackage{dcolumn}
\usepackage{bm}
\usepackage{color}

\begin{document}

\title{Planar Josephson junctions for sensors and electronics:Different geometry, new functionality.
}


\author{Vladimir M. Krasnov } 


\affiliation{
           Department of Physics, Stockholm University, AlbaNova University Center, 10691 Stockholm, Sweden. \\
              \email{vladimir.krasnov@fysik.su.se}           
}


\begin{abstract} 
Josephson junctions are key elements in superconducting electronics. The most common type is the overlap (sandwich-type) junction, formed by vertically stacking two superconducting layers. In contrast, planar junctions are fabricated without overlap, at the edge of two superconducting films within a single plane. This geometric distinction has a significant impact on their physical properties.
The planar geometry greatly enhances sensitivity to magnetic fields and improves impedance matching for terahertz (THz) devices. Its two-dimensional structure allows for simple and flexible electronic component design, enabling drastic miniaturization. Here I highlight recent advances in the application of planar junctions for novel technologies, including junction-on-cantilever sensors for super-resolution magnetic imaging, vortex-based memory cells, and programmable superconducting diodes. I will also discuss the general requirements, future perspectives, and key challenges in the evolving field of superconducting electronics.

\end{abstract}

\maketitle

\section*{Introduction}
\label{intro}

We live in the time of a third industrial (digital) revolution. It is characterized by an exponential growth of the data volume, which is expected to reach 200 ZB this year. 
The explosive digitalization sets several challenges for the sustainable development of our society. According to IEA executive summary 2024 \cite{IEA}:
``Electricity consumption from data centres, artificial intelligence and the cryptocurrency sector could double by 2026. Data centres are significant drivers of growth in electricity demand in many regions... 
Updated regulations and technological improvements, including on efficiency, will be crucial to moderate the surge in energy consumption from data centres." 

At the same time, modern semiconducting electronics is reaching its physical limits on many fronts (memory-, heat-, power-walls, interconnect bottleneck, etc.) \cite{Limits}. 
The ``golden age" of semiconducting electronics, when it was possible to improve performance by just miniaturization has ended approximately three decades ago. For the last two decades the development of semiconducting computation has been propelled by the expansion into the third dimension (e g. 3D FinFET) and changes in computer architecture \cite{Computing_2017}. 
However, even this approach shows a tendency for saturation. Therefore, radical innovative solutions are needed to maintain the present digitalization trends. 

The growing demands for digital data capacity have led to renewed interest in the development of a classical superconducting computer \cite{Holmes_2013,Ortlep_2014,Golod_2015}. It could enable drastic gains in both speed and power efficiency compared with semiconducting analogs.  
However, the existing rapid single flux quantum (RSFQ) concept of a digital superconducting computer \cite{Ortlep_2014,Likharev_1991,Tolpygo_2016,Takeuchi_2015,Mukhanov_2019,Semenov_2017,Tolpygo_2019} has met a ``scalability" wall due to the limits of miniaturization of its key component, the superconducting quantum interference device (SQUID) \cite{Golod_2015,Tolpygo_2016}.  
The problem is associated with the need of large enough inductance and critical current to store the flux quantum (the SQUID parameter $\beta_L \geq 3$). Therefore, innovative solutions, aiming at replacement of SQUIDs by scalable to submicron sizes superconducting components, are needed for the development of future generation of computers. Numerous new ideas on how to replace SQUIDs have been generated in the past decade \cite{Golod_2015,Soloviev_2017,Golod_2023,Alam_2023} and await for experimental scrutiny. 

Geometry plays a crucial role in defining the physical properties and operational performance of superconducting and spintronic components. Contemporary superconducting electronics is dominated by overlap-type Josephson junctions (JJs). These junctions are fabricated through the sequential deposition of superconducting and interlayer films, with subsequent interface modifications (e.g., oxidation). This approach enables precise control over interlayer thickness, composition, and interface transparency, which is particularly important for tunnel JJs.

Highly reproducible fabrication processes have been developed for various types of overlap JJs, such as Nb/AlAlO$_x$/Nb tunnel junctions~\cite{Semenov_2017,Tolpygo_2019} and Nb/Nb$_x$Si$_{1-x}$/Nb proximity-coupled junctions~\cite{Benz_2011}, with the total number of junctions on a single chip approaching one million~\cite{Semenov_2017}. The relative simplicity of the fabrication method—compatible with standard microfabrication tools—along with the robustness and wide tunability of these junctions, makes overlap JJs the preferred choice for practical applications today.

Planar~\cite{Note} junctions are formed without overlapping superconducting electrodes, as sketched in Fig.\ref{fig:1}(a). Various types of planar JJs have been studied over the years, including proximity-coupled planar JJs and variable-thickness-bridge JJs \cite{Likharev_1979}, grain-boundary JJs in high-$T_c$ cuprates~\cite{Kirtley_1996,Boris_2013}, junctions based on two-dimensional electronic systems in semiconducting heterostructures~\cite{Krasnov_2005a,Baumgartner_2022}, van der Waals materials~\cite{Kim_2021,Herrero_2021,Zhang_2020}, topological insulators~\cite{Kudriashov_2025}, and more.

Different geometries are associated with different physical properties and new functionalities. As Likharev noted in 1979~\cite{Likharev_1979}, “As regards applications, it is generally felt that variable-thickness bridges, given a proper choice of materials for the span and the banks, may meet almost all the practical requirements.” However, planar JJs have not yet seen widespread use in superconducting electronics, largely due to the lack of simple and reliable fabrication procedures.

The development of focused ion beam (FIB) techniques has opened new possibilities for the direct writing of planar JJs~\cite{Tarte_1999,Krasnov_2005,Golod_2010,Cox_2014}. Figure~\ref{fig:1}(b) shows a schematic of the fabrication process using Ga$^{+}$ FIB. A planar junction can be made from either a double-layer structure or a single superconducting film. In the double-layer case, milling through the top superconducting layer yields proximity-coupled JJs—of the SNS type~\cite{Tarte_1999,Golod_2021,Aarts_2023} if the bottom layer is a normal metal (N), or of the SFS type~\cite{Golod_2015,Krasnov_2005,Golod_2010,Golod_2021,Aarts_2017,GolodAV_2019,GolodSFS_2019,Golod_2022} if it is a ferromagnet (F). Making a trench in a single superconducting film results in a variable-thickness-bridge type JJ~\cite{Golod_2022,Grebenchuk_2022}. He$^+$ FIB has also been successfully employed to fabricate damage-induced planar JJs in YBa$_2$Cu$_3$O$_{7-\delta}$ high-$T_c$ films~\cite{Cybart_2022,Goldobin_2024}. The high reproducibility of FIB-fabricated planar JJs has enabled advanced prototyping of novel devices, revealing numerous advantages and promising prospects for the use of planar JJs in superconducting electronics.

This work provides an overview of the advantages associated with planar geometry, as well as recent developments and perspectives on superconducting components based on planar JJs.

\section*{Planar geometry: differences and advantages}
\label{sec:1}

\begin{figure*}
\includegraphics[width=0.99\textwidth]{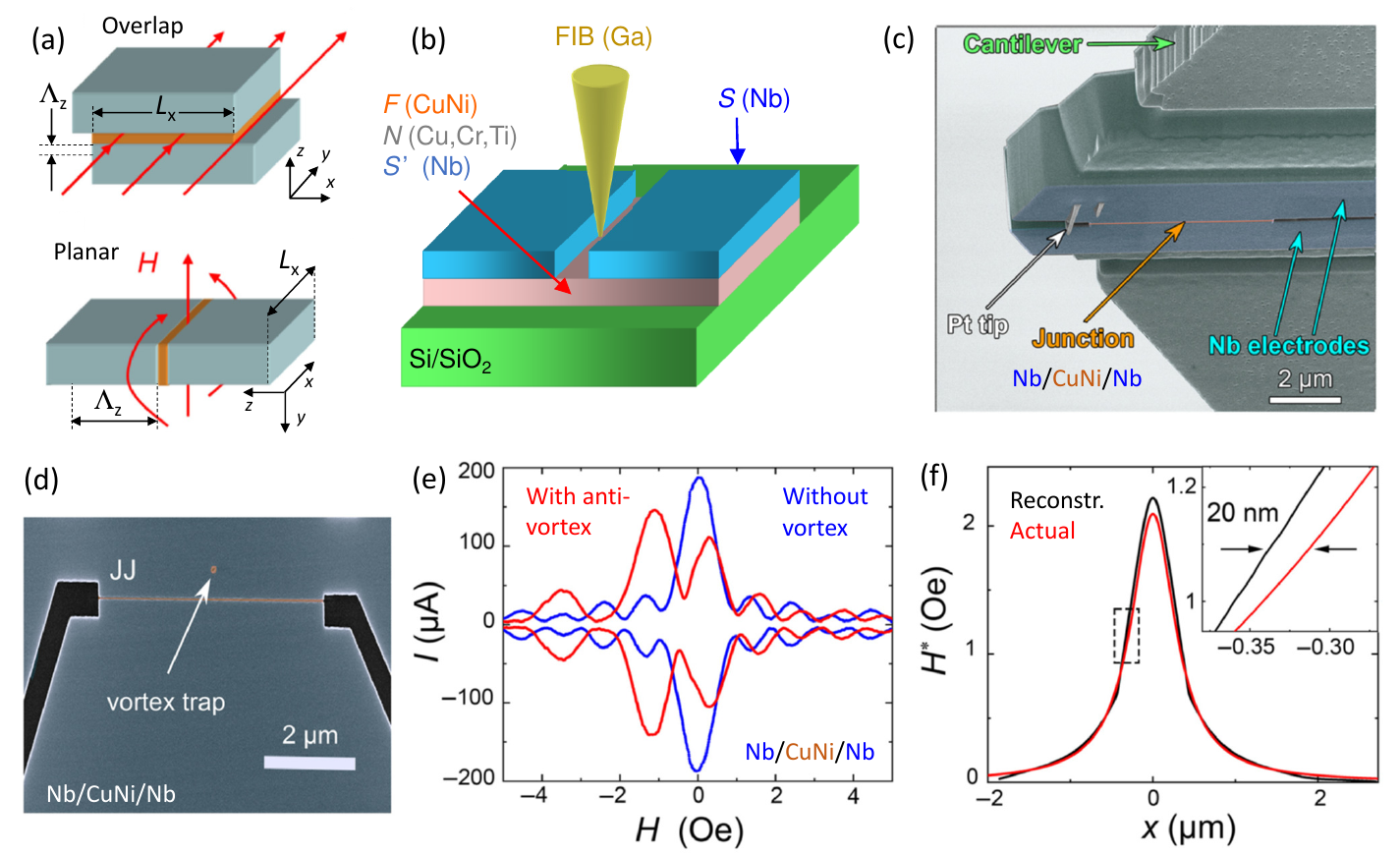}
\caption{(a) A sketch of geometry difference between overlap and planar Josephson junctions. 
(b) A sketch of planar junction fabrication by focused ion beam. 
(c) SEM image of a scanning probe sensor with a Nb/CuNi/Nb planar junction on a cantilever. 
(d) SEM image of a planar Nb/CuNi/Nb junction with a vortex trap in one electrode. 
(e) measured $I_c(H)$ characteristics of the junction from (d) in the Meissner state (blue) and with a trapped antivortex (red). 
(f) Holographic reconstruction of vortex stray field in the junction using the distorted (red) $I_c(H)$ pattern from (e) as a hologram. The inset shows the difference between the actual and reconstructed field profiles. The accuracy of reconstruction $\sim 20$ nm is much smaller than the junction length $L_x~\sim 5~\mu$m, which indicates super-resolution imaging. Panels (a), (b) and (c-f) were adopted from Refs. \cite{Boris_2013}, \cite{Krasnov_2005} and  \cite{Hovhannisyan_2023}, respectively.
}
\label{fig:1}       
\end{figure*}

Planar JJs have a geometry that is orthogonal to that of overlap JJs, see the orientations of ($xy$) junction planes in Fig. \ref{fig:1} (a).
While this may seem like a purely technical distinction, it leads to profound qualitative and quantitative differences between planar and overlap JJs.
These differences offer new functionalities and open up possibilities for novel applications.

The list of differences and advances of planar JJs include:

\paragraph*{I. Openness} (difference).

Planar geometry opens up the junction cross-section for detailed inspection.
As noted by Likharev, "Naturally, variable-thickness bridges and similar structures are quite convenient in physical research into the processes occurring in weak links" \cite{Likharev_1979}. For example, this allows direct imaging of Josephson vortices by means of scanning probe microscopy \cite{Roditchev_2015,Dremov_2019,Chen_2024}. 

The openness, however, does not deteriorate the stability of the junctions. They remain remarkably stable over time, with no significant variations observed over a period of 10 years in air, without encapsulation or any special protection. Furthermore, we tested annealing of the JJs for a few hours in air at $T=300^{\circ}$~C, which likewise did not lead to any noticeable parameter drift. Most likely, the chemical stability is due to the formation of a robust, protective Nb-oxide layer.

\paragraph*{II. Non-local electrodynamics} (difference).

Overlap JJs have local electrodynamics, described by the sine-Gordon equation \cite{Barone}. It is local in the sense that the Josephson current density at a given point, $J(x)$, is determined by the local Josephson phase difference, $\varphi(x)$, via the dc-Josephson (current-phase) relation. For tunnel JJs, $J(x)=J_c(x)\sin[\varphi(x)]$. Electrodynamics of planar JJs is non-local because the Josephson current is determined by the integral of phases over the whole junction length \cite{Likharev_1979,Mintz_2001,Clem_2010}. This has a profound effect on JJ characteristics \cite{Boris_2013}. For example, it leads to unusual temperature dependencies with non-vanishing lower critical field, $H_{c1}$, for Josephson vortices, and non-diverging Josephson penetration depth, $\lambda_J$, at $T\rightarrow T_c$.

\paragraph*{III. Large demagnetization factor} (difference).

Planar junctions are sensitive to out-of-plane magnetic field, for which superconducting electrodes have a large demagnetization factor, $n_y$. For long and thin electrodes, $L_z\gg L_x\gg d$, $n_y\simeq 1 -d/L_x$ \cite{Osborn_1945}. The effective magnetic field at the edge of the electrode is,
\begin{equation}
H_{eff}=\frac{H_y}{1-n_y}\simeq \frac{L_x}{d}H_y \gg H_y.
\label{Heff}
\end{equation}
For JJs with $L_x=5~\mu$m and $d=70$ nm, $H_{eff} \simeq 71~H_y$.

\paragraph*{IV. High sensitivity to magnetic field} (advantage).

Flux quantization in JJ leads to Fraunhofer-like modulation of critical current, $I_c(H)$, in external magnetic field, $H$. Magnetic field sensitivity, $dI_c/dH \sim I_{c0}/\Delta H$, is determined by the flux-quantization field, $\Delta H=\Phi_0/L_x\Lambda_z$, where $L_x$ is the JJ length and $\Lambda_z$ is the effective magnetic thickness of the JJ. For overlap JJs, $\Lambda_z = t+\lambda_{L1}\tanh(d_1/2\lambda_{L1})+\lambda_{L2}\tanh(d_2/2\lambda_{L2})$, where $t$ is the junction interlayer thickness, $d_{1,2}$ are thicknesses, and $\lambda_{L1,2}$ are London penetration depths of the two superconducting electrodes \cite{Barone}. The maximum value of $\Lambda_z$ for large $d_{1,2}$ is $\simeq 2\lambda_L$. Thus, an overlap JJ with $L_x=5~\mu$m and $\lambda_L=100$ nm (typical for sputtered Nb) would have $\Delta H\simeq 10$ Oe.

For planar JJs, the perpendicular magnetic field cannot pinch through the electrodes and is guided along the electrodes towards the edges, leading to the large demagnetization factor. For short electrodes, $L_z < L_x$, approximately half of the total incoming flux, $\Phi=H_yL_xL_z$, goes through the JJ and the rest closes on the opposite side of the electrode. For long electrodes, $L_z>L_x$, the length scale for flux penetrating into the junction is determined by $L_x$. This leads to quantization fields \cite{GolodSFS_2019,Clem_2010},
\begin{eqnarray}
\Delta H \simeq 1.8 \frac{\Phi_0}{L_x^2}, ~~~~ L_{z1,2}\gg L_x,\\
\Delta H \simeq 2\frac{\Phi_0}{L_x(L_{z1}+L_{z2})}, ~~~~ L_{z1,2}\ll L_x,\\
\Delta H \simeq 2 \frac{\Phi_0}{L_x(L_x/1.8 + L_{z2})}, ~~~~ L_{z1}>L_x>L_{z2}.
\end{eqnarray}
Thus, the effective magnetic thickness of planar JJs is determined by the size of electrodes, rather than London penetration depth. For a planar JJ with $L_x= 5~\mu$m, $\Delta H \sim 1$ Oe, is almost an order of magnitude smaller than for similar-size overlap JJs. Note that the enhancement of flux sensitivity is smaller than the edge field enhancement, $(1-n_y)^{-1}$. This happens because demagnetization fields are non-uniform with maxima at the edges and zero in the middle of electrodes.

The described above flux-focusing phenomenon in planar JJs leads to a drastic enhancement of magnetic field sensitivity. As noted in Ref. \cite{GolodSFS_2019}, SFS-type planar JJs have sensitivities comparable to similar-size superconducting quantum interference devices (SQUIDs). Furthermore, planar JJs could enable super-resolution magnetic imaging with spatial resolution not limited by the size of the sensor \cite{GolodSFS_2019,Hovhannisyan_2023}.

Fig. \ref{fig:1} (c) shows a scanning electron microscope (SEM) image of a scanning probe sensor with a Nb/CuNi/Nb planar junction on a cantilever \cite{Hovhannisyan_2023}. Fig. \ref{fig:1} (f) demonstrates an example of super-resolution reconstruction of stray magnetic fields from an Abrikosov vortex (AV). The inset indicates that the spatial accuracy of reconstruction is $\sim 20$ nm, much smaller than the junction size $\sim 5~\mu$m. Thus, planar JJ sensors allow obviation of a tradeoff problem between spatial resolution and magnetic field sensitivity in scanning probe sensors \cite{GolodSFS_2019,Hovhannisyan_2023}.

\paragraph*{V. Abrikosov vortices: manipulation, phase-shift and readout} (advantages).

\begin{figure*}
\includegraphics[width=0.95\textwidth]{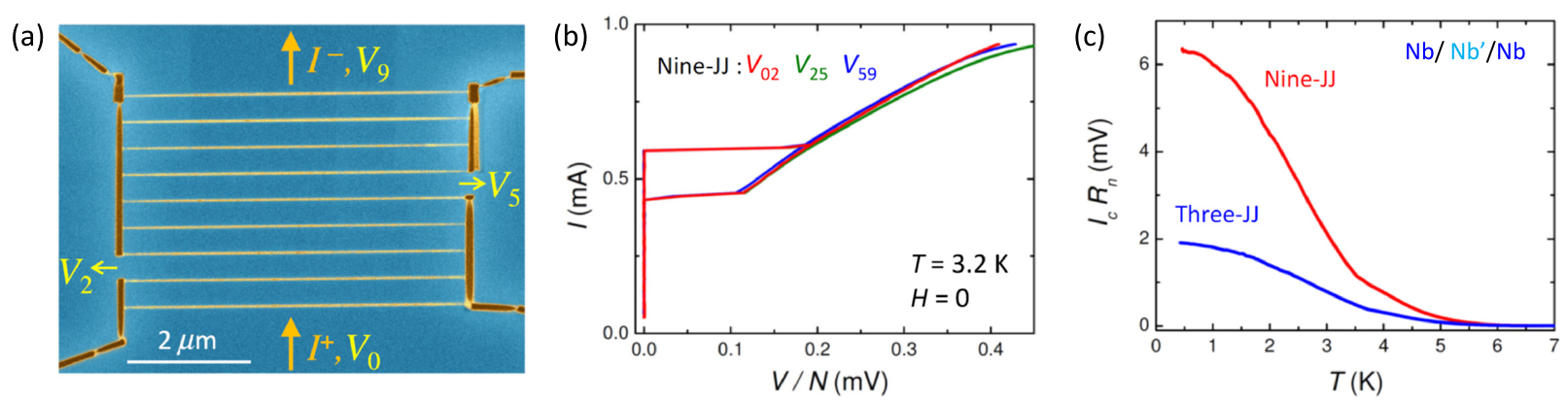}
\caption{(a) SEM image (false color) of an array with nine variable-thickness bridge type planar Nb junctions, and with additional electrodes to intermediate electrodes 2 and 5. (b) Current - Voltage characteristics of the three array segments, normalized by the number of JJs in the segment. Coincidence of the curves indicates the reproducibility of the fabrication technique and perfect current-locking of all JJs. (c) Temperature dependence of $I_c R_n$ for two arrays with three and nine JJs. 
Adopted from Ref. \cite{Grebenchuk_2022}.}
\label{fig:2}       
\end{figure*}

Abrikosov vortices in thin superconducting films are oriented so that currents are circulating in-plane and the magnetic field is out-of-plane. A transport current with density $J$ exposes a Lorentz force on AV, 
\begin{equation}
    F_L= d [J\times\Phi_0],
    \label{F_L}
\end{equation}
where $d$ is the film thickness.

For overlap JJs, vortices create problems. The vortex pinches through the junction plane, and the current is co-aligned with the vortex flux (both in the $z$-direction in Fig. \ref{fig:1} a), leading to a vanishing Lorentz force. Therefore, it is difficult to control and manipulate AVs in overlap JJs \cite{Lisitskii_1992,Finnemore_1994}.

For planar JJs, AVs are also perpendicular to the film (in the $y$-axis). However, they are parallel to the JJ and cannot cross it \cite{Golod_2010}. The transport current is flowing along the electrode ($z$-axis), leading to a Lorentz force parallel to the JJ (in the $x$-direction) across the electrode. This enables simple control and manipulation of AVs. Therefore, AVs in planar junctions represent not a problem, but an opportunity.

Stray flux and circulating currents from AV induce a geometry-dependent Josephson phase shift in a planar JJ \cite{GolodAV_2019}. The exact position of the vortex can be defined with the help of artificial vortex traps \cite{Golod_2015,Golod_2010,GolodAV_2019}, as illustrated in Fig. \ref{fig:1} (d). If AV is placed close to the JJ, approximately half of its flux, $\Phi_0/2$, goes into the JJ, leading to formation of switchable $0-\pi$ junctions \cite{Golod_2010,Goldobin_2007} with a $\pi$ phase shift within the JJ. The phase shift is geometrically tunable \cite{GolodAV_2019}. By varying junction length and vortex trap position it is possible to create phase shifted $0-\varphi$ JJs with arbitrary phase shift $\varphi \in [-2\pi, 2\pi]$ with the help of a single vortex or antivortex \cite{GolodAV_2019}. In a similar manner, AV could create a flux offset in a SQUID loop \cite{Golod_2021}. 

The vortex-induced phase shift in planar JJs causes a distortion of $I_c(H)$ patterns \cite{Golod_2010,GolodAV_2019}, as illustrated in Fig. \ref{fig:1} (e). In the absence of the vortex (blue lines) the pattern is Fraunhofer-like with a central maximum at $H=0$ (although it is not completely periodic due to non-locality of planar JJs \cite{Clem_2010}). 
For a $0-\pi$ junction, the central maximum is replaced by a minimum \cite{Golod_2010,Goldobin_2007}, as can be seen from the red lines in Fig. \ref{fig:1} (e). This leads to a maximum distinction between Meissner (maximal $I_c$) and vortex (minimal $I_c$) states at zero magnetic field \cite{Golod_2015}. 

The ease of vortex manipulation and the large distinction between vortex states make planar junctions suitable for storing digital information in the form of a single AV \cite{Golod_2015,Golod_2023}. Since vortex manipulation (write/erase operations) is performed at the AV depinning current, whereas state readout is performed at the Josephson critical current (which is typically much smaller) the readout of vortex states can be carried out non-destructively \cite{Golod_2015,Golod_2010,Skog_2024}.

\paragraph*{VI. Flexibility of geometrical design} (advantage).

\begin{figure*}
  \includegraphics[width=0.95\textwidth]{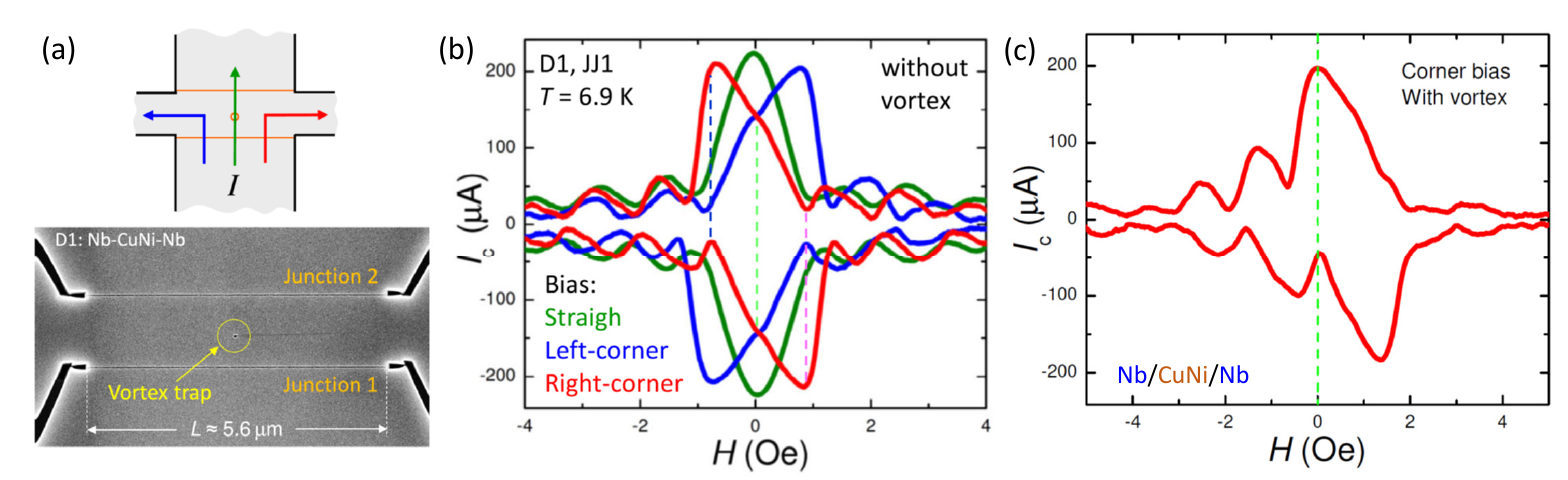}
\caption{A sketch (top) and a SEM image (bottom) of a four-terminal Josephson diode with a vortex trap. (b) Magnetic field modulation of critical current in the absence of vortex for three bias configurations sketched in (a). The asymmetric corner bias leads to a profound nonreciprocity at finite magnetic field. (c) $I_c(H)$ modulation for the right-corner bias with a trapped vortex. Note that the largest nonreciprocity is shifted to zero field. The diode polarity of such a diode is switchable by changing the sign of trapped vortex and the bias configuration. Adopted from Ref. \cite{Golod_2022}.}
\label{fig:3}       
\end{figure*}

The openness of the planar layout allows a great deal of design and fabrication flexibility \cite{Cox_2014}, enabling effective utilization of geometric advantages. For instance, it is not difficult to fabricate JJs of arbitrary shape with specifically designed nonuniform distributions of critical current density and electrode width along the junction, or to realize various geometrical constructions that are difficult or impossible to achieve with overlap geometry.

The flexibility of the planar geometry opens up a variety of new possibilities. For example, in JJ arrays, contacts to intermediate electrodes are very useful for studying the intricate dynamics of individual JJs during array synchronization, as well as for probing defects, nonuniformities, and the spread of junction characteristics. In overlap-type stacked JJs, it is technically challenging to make contacts to intermediate electrodes, as the need for insulating layers rapidly increases the structural height and unevenness. This issue does not arise in planar JJ arrays.

Figure~\ref{fig:2} (a) shows a SEM image of an array of nine variable-thickness planar Nb JJs with contacts to intermediate electrodes $\#2$ and $\#5$ \cite{Grebenchuk_2022}. Thus, the characteristics of array segments comprising JJs 1+2, 3+4+5, and 6+7+8+9 could be measured simultaneously. Figure~\ref{fig:2} (b) presents the corresponding current-voltage ($I$--$V$) characteristics, with voltage normalized by the number of JJs in each segment. The matching of normal resistances indicates the reproducibility of the fabrication process, resulting in nearly identical JJs. The coincidence of switching currents demonstrates perfect current locking of the array. Simultaneous switching of all JJs leads to a multiplication of the readout voltage, exceeding 6 mV at low temperatures, as shown by the red line in Fig.\ref{fig:2} (c). Such cascade multiplication of the readout voltage in JJ arrays may be useful for microwave (MW) and THz detectors \cite{Cattaneo_2025}.

Figure~\ref{fig:3} (a) shows an example of a superconducting diode with switchable polarity and nonreciprocity. It is based on four-terminal Nb/CuNi/Nb planar JJs \cite{Golod_2022}. Nonreciprocity of the critical currents requires simultaneous breaking of both spatial and time-reversal symmetries. In this device, spatial symmetry can be broken controllably and tunably by selecting different bias configurations.

Figure~\ref{fig:3} (b) presents measured $I_c(H)$ patterns for three bias configurations, as sketched in panel (a). The JJ behaves reciprocally in the symmetric bias configuration (olive), while for asymmetric bias a strong nonreciprocity, $A = \left| I_c^+ / I_c^- \right| \simeq 10$, is achieved at a finite magnetic field, which breaks time-reversal symmetry. Time-reversal symmetry can also be broken by the introduction of AV. The device includes a vortex trap, marked in Fig. \ref{fig:3} (a). Fig. \ref{fig:3} (c) shows $I_c(H)$ modulation with a trapped AV in the right-corner bias configuration. In this case, nonreciprocity of $A \simeq 4$ is observed at zero magnetic field, as indicated by the vertical dashed line.

The diode is one of the fundamental components in electronics, with a wide range of applications in both analog and digital circuits. Diodes based on other types of planar JJs have also been demonstrated \cite{Baumgartner_2022,Zhang_2020,Goldobin_2024}. Developing a technologically simple and highly efficient superconducting diode will be indispensable for next-generation classical superconducting computers and high-frequency signal processing.

However, the demonstrated planar JJ diode is not merely an efficient diode—its most remarkable feature is reconfigurability. The diode polarity can be easily reversed by changing the bias configuration and the polarity of the trapped vortex \cite{Golod_2022}. This opens up new possibilities for creating novel components with advanced functionality, such as programmable logic gates.

\paragraph*{VII. Terahertz applications: impedance matching and synchronization} (advantage).

\begin{figure*}
  \includegraphics[width=0.95\textwidth]{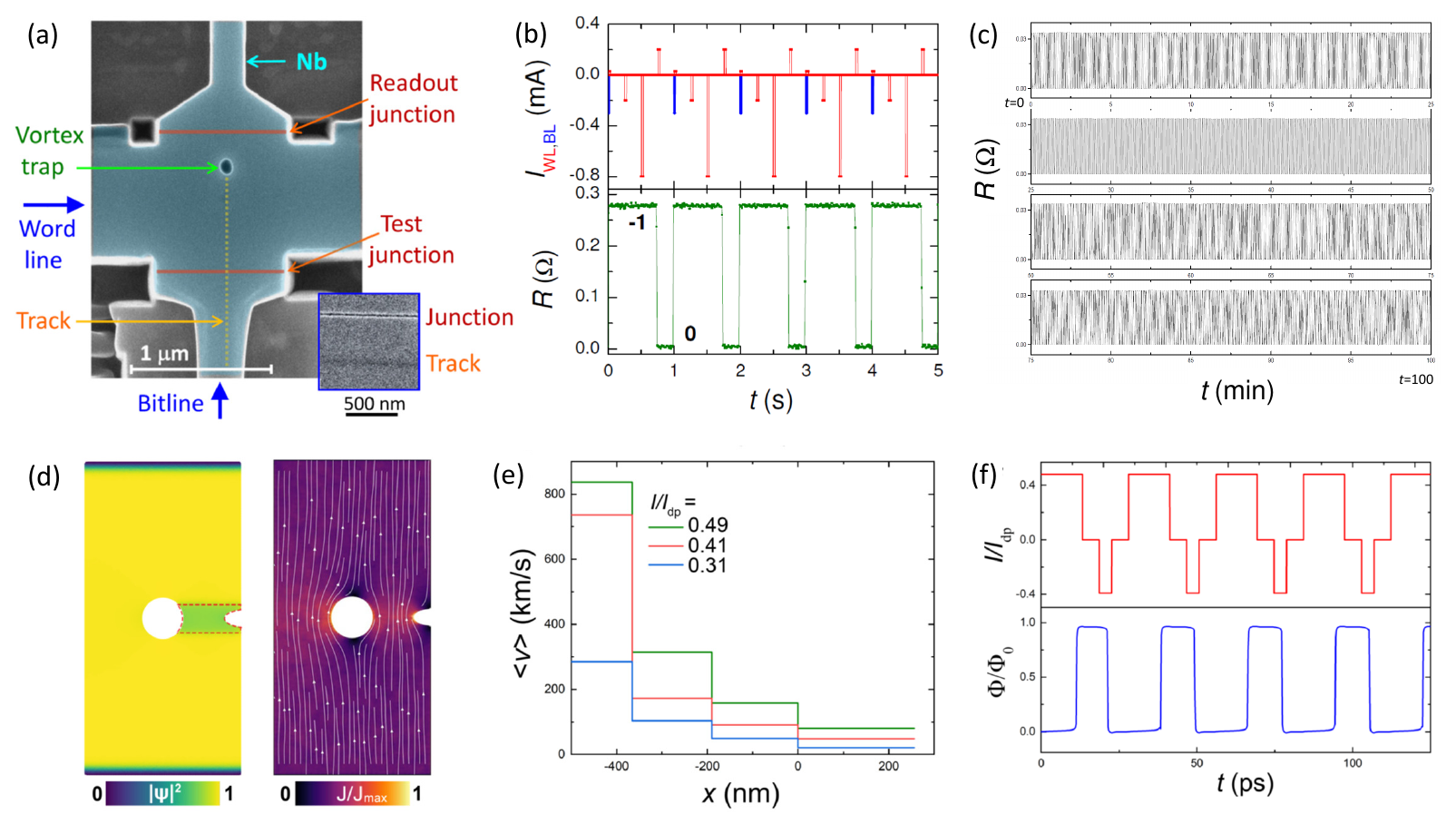}
\caption{(a) SEM image (false color) of a $\sim 1\times 1~\mu$m$^2$ vortex-based AVRAM memory cell. (b) Demonstration of write and erase operation by word-line (red) and bit-line (blue) current pulses. The top panel shows current pulse sequence. The bottom panel represents the resistance of readout JJ. (c) Demonstration of a long-time fault free operation for another device. (d-f) TDGL modeling of cell operation. (d) Cell geometry with spatial distribution of the order parameter (left) and current density (right). The width of the electrode is $1~\mu$m, as in (a). (e) Average vortex velocities along the path from the right to the left edge at three bias currents. (f) Demonstration of ultrafast write/erase operation by bipolar current pulses. Panels (a,b), (c) and (d-f) are adopted from Refs. \cite{Golod_2023}, \cite{Golod_2015} and \cite{Skog_2024}, respectively. }
\label{fig:4}       
\end{figure*}

Superconducting electronics can operate in the THz frequency range \cite{Borodianskyi_2017}, which is crucial for future ultrafast electronics and telecommunication systems \cite{Kakeya_2024}. The upper frequency limit of a JJ is determined by its characteristic voltage, $V_c = I_c R_n$ \cite{Barone}, where $R_n$ is the normal-state resistance of the JJ. Constriction-type JJs, such as variable-thickness bridge, are characterized by a large $V_c$, which can exceed that of tunnel JJs by up to a factor of two \cite{Kulik_1977}. FIB-patterned Nb variable-thickness JJs, shown in Fig.~\ref{fig:2} (c), reach relatively high values of $V_c \simeq 0.8$~mV at low temperatures. The ultimate theoretical limit, $V_c = \pi \Delta / e$, where $\Delta$ is the superconducting energy gap, is achieved in ballistic (clean-limit) constriction JJs \cite{Kulik_1977}. For Nb, this corresponds to a maximum $V_c \sim 5$~mV, implying that such JJs could operate at frequencies up to $f \sim V_c / \Phi_0 \sim 2.5$~THz.

Efficient transmission and absorption of THz signals require proper impedance matching of circuit components. The impedance of effective THz devices should be comparable to a fraction of the free-space impedance. A JJ by itself cannot absorb or emit THz radiation efficiently because its physical dimensions are much smaller than the radiation wavelength,
$\lambda_0 = 300~\mu$m at $f = 1$ THz,
leading to a significant impedance mismatch with free space \cite{Krasnov_2023}. The planar geometry offers a straightforward solution to this problem. For example, a planar JJ, sketched in Fig. \ref{fig:1} (a), can function as a standard dipole antenna if the electrode lengths are $L_z = \lambda_0^*/4$ (where $\lambda_0^* < \lambda_0$ accounts for the substrate’s dielectric constant).
In this configuration, the electrodes act as THz antennas, and their shapes and sizes can be readily tailored to specific application requirements, thanks to the design flexibility of the planar geometry.

JJ arrays offer further advantages for both emission \cite{Borodianskyi_2017,Kakeya_2024} and detection \cite{Cattaneo_2025} of THz radiation. However, achieving this requires coherent operation of the junctions. Synchronization of many JJs in a large array depends on the presence of long-range mutual interactions between them. As shown in Ref.\cite{Grebenchuk_2022}, planar JJ arrays—such as the one in Fig. \ref{fig:2} (a)—exhibit long-range coupling via stray magnetic fields, which cannot decay easily because they cannot penetrate the superconducting electrodes. Therefore, planar JJ arrays have good prospects for the development of coherent THz emitters and detectors.

\paragraph*{VIII. Compactness and scalability} (advantages).

Due to the absence of overlap, planar JJs have a very small lateral area. This is conceptually similar to the three-dimensional FinFET design in semiconductor electronics, where vertical expansion reduces the on-wafer footprint of components. In the array shown in Fig. \ref{fig:2} (a), the junction cut is approximately 30 nm wide, and the electrode width is around 500 nm—dimensions that can be readily reduced to 200 nm or even smaller (down to $\lambda_L\sim$100~nm). This enables the fabrication of compact devices with a high junction density.

Figure~\ref{fig:4} (a) shows a SEM image of the smallest vortex-based (AVRAM) memory cell demonstrated to date \cite{Golod_2023}. The device includes a $\sim 1 \times 1~\mu$m$^2$ Nb island, word-line and bit-line electrodes, a vortex trap, a guiding track for vortices, and two readout JJs (Nb variable-thickness bridges). Fig.~\ref{fig:4} (b) illustrates controllable write and erase operations performed using current pulses $I_{WL}$ and $I_{BL}$ applied to the word-line and bit-line, respectively. Figure~\ref{fig:4} (c) demonstrates robust, fault-free operation over extended period \cite{Golod_2015}.

The lack of dense RAM is one of the main bottlenecks for the development of digital superconducting electronics \cite{Ortlep_2014,Alam_2023}. The state-of-the-art RSFQ cells reported in Ref. \cite{Ortlep_2014} have a large size $60 \times 60~\mu$m$^2$. The AVRAM cell from Fig. \ref{fig:4} (a) is much smaller, highlighting a major advantage of the planar geometry in terms of device miniaturization. In fact, the entire AVRAM cell occupies a much smaller area than a single overlap JJ in RSFQ cell from Ref.~\cite{Ortlep_2014}. Multiple factors contribute to this advantage \cite{Golod_2015,Golod_2023}, with the compactness of planar JJs being just one. 

The maximum distinctiveness, the robust nondestructive readout of vortex states by planar JJs are another important advantages for memory applications. The RSFQ memory is based on storing of a flux quantum in a SQUID loop. In contrast to planar JJs, 0/1 states of a SQUID are non-distinguishable and the readout is destructive. Therefore, a second SQUID loop is needed for the nondestructive readout \cite{Ortlep_2014}. This significantly complicates the cell architecture and increases the RSFQ memory cell size. 
Furthermore, SQUIDs need large inductances to store magnetic flux, whereas the AVRAM cell does not. All these factors, together with the design flexibility inherent to the planar architecture lead to an extremely simple AVRAM cell structure compared to RSFQ, further contributing to its compactness. Combined with the nanometer-scale size of AV, this allows for sub-micron scalability of AVRAM technology.

\section*{Discussion}
\label{sec:2}

Although superconducting electronics could offer significant advantages, there are serious limitations and challenges. Very complex projects, such as the development of superconducting high-end computer, would require massive investments. This could happen only when the technology will reach high enough readiness level with demonstrated impressive characteristics.    

\subsection*{Target characteristics for the next generation superconducting electronics}

\paragraph*{Compatibility with RSFQ technology.}

The RSFQ concept \cite{Likharev_1991} is based on generation and utilization of quantized voltage pulses upon exit or entrance of a flux quantum in a SQUID loop. The RSFQ pulses are transferred via superconducting transmission lines at a fraction of the speed of light, $c^*\sim 1 \times 10^8$ m/s. Together with the sub-ps switching time of JJs in a SQUID, this enables an ultrafast operation. 

A major problem for the development of any type of ultrahigh-frequency electronics is the sensitivity to delay times. The conventional synchronous operation fails when the time of propagation of signals along the chip becomes comparable to the clock period. Therefore, asynchronous or delay insensitive electronics is required at high frequencies. An important advantage of RSFQ is that it enables asynchronous operation due to an inbuilt non-volatile memory. After operation, a flux state is preserved in the SQUID loop so that the next operation could be performed at any time within the clock period \cite{Likharev_1991}. 

RSFQ is the only fully operational concept of superconducting digital electronics today. Fairly complex RSFQ prototypes have been demonstrated \cite{RSFQ_2016,SFQ_100} and numerous smart innovative solutions were developed \cite{Likharev_1991}. It doesn't make sense to start from scratch. Although SQUIDs have to be replaced in components requiring very large scale integration (VLSI), such as RAM and logic gates, SQUIDs may be preserved in the rest of components. The next generation of superconducting digital electronics should be compatible with RSFQ and should built upon the large accumulated knowledge. 

\paragraph*{Operation frequency.} 
The future DDR6 generation of semiconducting electronics is expected to have the clock frequency of 6.4-8.5 GHz.
In order to impress, superconducting electronics should offer at least an order of magnitude improvement. That implies the base clock frequency of 64-85 GHz. 
These values are well within the achievable range of Nb-based superconducting electronics, which is able to operate above 1 THz \cite{Karpov_2007}. Thus, the near-future target value for the clock frequency of superconducting electronics is $\sim 100$ GHz, which is feasible for RSFQ \cite{SFQ_100}.  

The actual advance of superconducting computer in terms of operation speed is determined not only by the clock frequency but by the intrinsically fast (ps) switching times \cite{Likharev_1991}, as compared to 1-10 ns $RC$ times of semiconductor components. The long switching time leads to a long latency, strobe and charging times of semiconducting electronics. Therefore, the actual cycle time in the latest DDR5 DRAM (typical parameters 40-40-40-77) is well above hundred clock periods, bringing the actual operation speed back to 10 ns (100 MHz) range. 
Therefore, even at the same clock frequency, superconducting electronics would operate much faster than semiconducting electronics because it could have much shorter latency times. 

\paragraph*{Miniaturization.} 
Integration at least to the VLSI level is necessary for building fast and powerful computers. There are many well-known reasons and I will dwell on just one: shorter distances lead to smaller delay times and enable higher operation frequency and faster operation. The problem with delay times is rapidly escalating with increasing clock frequency. For 200 GHz the period is $\tau=5$ ps. For $c*=10^8$ m/s the $0.1~\tau=0.5$ ps delay is gained at a distance of $50~\mu$m, which is shorter than the RSFQ memory cell size from Ref. \cite{Ortlep_2014}.   
Therefore, scaling to sub-micron sizes is necessary for building (any type of) ultrafast computer. 

\paragraph*{Delay-insensitive electronics}
At high frequencies the delay problem becomes so acute that 
miniaturization alone can not solve it. Development of delay-insensitive components and architecture would be needed.

\paragraph*{Energy efficiency}
Although superconductors dissipate little power, they have poor thermal conductivity and are prone to self-heating. Therefore, energy per operation should be below $1$ aJ \cite{Tolpygo_2016} and static energy consumption should be eliminated \cite{Takeuchi_2015,Mukhanov_2019}.  


The list above is incomplete. There are numerous other challenges, not the least with the computer architecture \cite{Likharev_1991,Fourie_2019}, which would need to be addressed. But they are beyond this overview because they are not related to the junction geometry.

\subsection*{Characteristics of planar AVRAM cell}

Planar AVRAM cell, Fig \ref{fig:4} (a), satisfy most of the criteria mentioned above.

\paragraph*{Compatibility with RSFQ technology.} 

The vortex-based electronics \cite{Golod_2015,Golod_2023} is potentially fully compatible with RSFQ. Indeed it also operates with the flux quantum and has an inbuilt memory functionality, enabling asynchronous operation. 

\paragraph*{Operation frequency.} 

There is a long-standing misconception that AVs are inherently slow, with typical propagation velocities on the order of $\sim 1$~km/s. In reality, AVs are (practically) massless electromagnetic entities that can propagate at a speed of light. However, their actual velocity is influenced by pinning effects and viscous damping due to quasiparticle losses in the vortex core. Near the depinning threshold, AV motion can indeed be slow. Yet, as the current—and thus the driving Lorentz force—increases, the AV velocity grows rapidly and in a super-linear fashion.

Figures~\ref{fig:4} (d–f) show the results of time-dependent Ginzburg–Landau (TDGL) simulations of vortex dynamics in an AVRAM cell with a width $L_x = 1~\mu$m \cite{Skog_2024}. Fig.~\ref{fig:4} (e) presents the vortex velocity as it traverses from the right to the left edge of the superconducting island. The velocity is seen to increase sharply, reaching nearly 1000~km/s at the exit (left edge). Notably, this increase is not due to acceleration (since the Lorentz force is constant and the AV has no mass), but rather stems from the intrinsic nonlinearity of flux-flow viscosity. As velocity increases, the vortex viscosity decreases, enabling ultrafast propagation. Combined with the small physical size of the cell, this results in picosecond-scale operation times.

Figure~\ref{fig:4}(f) shows simulated AV write/erase operations driven by short bipolar current pulses. The switching times—defined as the delay between the onset of the current pulse (top panel) and the change in magnetic flux in the trap (bottom panel)—are in the ps range. The switching time can be estimated as
$\Delta t =L_x/2\langle v \rangle$. 
Taking $\langle v \rangle = 200$ km/s and $L_x = 1~\mu$m yields $\Delta t = 2.5$~ps, which agrees well with the simulation results in Fig.\ref{fig:4} (f) \cite{Skog_2024}.

\paragraph*{Miniaturization.} 
AV is the smallest magnetic object in type-II superconductors with the largest size determined by $\lambda_L \simeq 100$ nm (Nb). Therefore, AVRAM cell should be miniaturizable to approximately $2-3\lambda_L \simeq 200-300$ nm. 
This is three-four order of magnitude smaller than state of the art RSFQ cells. 

\paragraph*{Energy efficiency}
Energy dissipation in AVRAM cell during write/erase operation is equal to the work done by the Lorentz force, Eq. (\ref{F_L}) \cite{Golod_2015,Skog_2024}, 
\begin{equation}
    E=\frac{\Phi_0 I }{2}.
\end{equation}
For the cell from Fig. \ref{fig:4} (a) with switching currents $I\simeq 200~\mu$A, Fig, \ref{fig:4} (b), $E\simeq 0.2$ aJ. 

As for the challenges facing planar devices, they are numerous.

First, while FIB patterning is well-suited for prototyping, it is not viable for large-scale production. Therefore, the development of a new, scalable, and straightforward fabrication technology for planar JJs with controllable and reproducible properties is essential.

To date, only relatively simple proof-of-concept planar devices have been demonstrated. In order to achieve a sufficient technology readiness level to attract investment and industrial interest, it is necessary to realize more complex circuits and functionalities. This includes demonstrations of RAM, logic gates, full adders, and other key digital components. Furthermore, high-frequency operation must be validated, and compatibility with established RSFQ technology, must be ensured.

\subsection*{Prospects and challenges of superconduction electronics.}

Even though superconducting electronics cannot compete with semiconductor technology in terms of miniaturization, it offers advanced functionalities that may enable competition in terms of functionality per unit area. For example, flux-based superconducting memory is both persistent and non-volatile, while also maintaining fast operation speeds. There is no semiconductor equivalent that combines these features; for instance, flash memory is persistent but slow with erase times in the millisecond range.

The switchable vortex-based superconducting diode demonstrated in Fig.~\ref{fig:3} opens new possibilities for implementing programmable logic, such as field-programmable gate arrays (FPGAs).

A perfect superconducting diode is indispensable for the future development of superconducting electronics. It addresses one of the major limitations in superconducting circuit design: the absence of effective switches for signal routing and isolation. In semiconductor electronics, a single field-effect transistor can open or close a line for signal transfer. In contrast, superconducting circuits lack such a switching element, resulting in substantial current leakage across interconnections. For DC bias currents, this crosstalk issue can only be mitigated by providing separate bias lines, which significantly increases architectural complexity in VLSI circuits—commonly referred to as the interconnect bottleneck.
Even for high-frequency RSFQ pulses, signal leakage remains a significant concern due to finite mutual inductance and the absence of a defined signal propagation direction—pulses in a superconducting transmission line can travel equally well in both directions. This necessitates additional design measures, such as introducing current gain or creating a gradient in switching currents along the signal propagation path \cite{Likharev_1991}. These challenges pose serious constraints for scaling up to VLSI levels. A perfect superconducting diode would serve as a crucial new passive element, enabling reliable signal and current routing in superconducting systems.

In conclusion, planar Josephson junctions differ from conventional overlap JJs in both geometry and physical properties. This distinction enables drastic miniaturization, flexible circuit design, and novel device functionalities, opening the door to a broad range of emerging applications. In this work, I highlighted examples including super-resolution magnetic field sensors, efficient THz oscillators and detectors, programmable superconducting diodes, and scalable vortex-based memory and logic. This list is by no means exhaustive but is intended to inspire further creative exploration into the unique advantages offered by planar geometries.


\begin{acknowledgements}
The work was supported by the Knut and Alice Wallenbergs Foundation, PoC grant Nr. 2024.0319. 

I am grateful to my colleagues T. Golod, R. Hovhannisyan, A. Skog, R. Cattaneo, S. Grebenchuk and L. Morlet-Decarnin, who participated in experimental works included in the adopted Figures. 
\end{acknowledgements}


\section*{Conflict of interest}
The author declares that he has no conflict of interest.


\begin{thebibliography}{}
\bibitem{Note}  The term ``planar" may sound confusing for people familiar with semiconducting electronics since there ``planar" junctions are formed along the substrate (as in overlap JJs), while non-planar junctions are expanding in the third (vertical) dimension, as in 3D FinFET. Using similar logics, the term ``planar junction" has been sometimes applied to overlap JJ. Also other names like ``lateral" or ``in-line" JJs could be found in literature, further adding to the confusion. It is not much we can do about it, but emphasizing that here the term ``planar" refers to JJs with two electrodes placed in one (substrate) plane $(xz)$. The junction cross-section is orthogonal to the current flow direction ($z$-axis). Therefore, the effective junction plane ($xz$) is perpendicular to  the substrate plane, see Fig. \ref{fig:1} (a). Such definition is valid even when junction channel is two-dimensional (e.g. a graphene sheet).
\end{thebibliography}


\end{document}